\documentclass{SciPost}

\hypersetup{
    colorlinks,
    linkcolor={red!50!black},
    citecolor={blue!50!black},
    urlcolor={blue!80!black}
}

\usepackage[bitstream-charter]{mathdesign}
\usepackage{subcaption}
\usepackage{fancyhdr}
\usepackage{caption}
\usepackage{lineno}
\usepackage[binary-units]{siunitx}
\usepackage[nolist, nohyperlinks]{acronym}
\begin{acronym}[TDMA]
\acro{AI}{artificial intelligence}
\acro{FP}{Fabry-Perot}
\acro{PDH}{Pound-Drever-Hall}
\acro{DoF}{degree of freedom}
\acro{RL}{reinforcement learning}
\acro{DRL}{deep reinforcement learning}
\acro{Meta-RL}{meta reinforcement learning}
\acro{ML}{machine learning}
\acro{MDP}{Markov decision process}
\acro{GW}{gravitational wave}
\acro{RT}{real-time}
\acro{DDPG}{deep deterministic policy gradient}
\acro{TD3}{Twin Delayed DDPG}
\acro{SAC}{Soft Actor Critic}
\acro{PPO}{Proximal Policy Optimization}
\acro{RNN}{recurrent neural network}
\acro{GRU}{gated recurrent unit}
\acro{DPG}{deterministic policy gradient}
\acro{DQN}{deep Q-network}
\acro{TDS}{time delay system}
\acro{ADC}{analogue-to-digital converter}
\acro{DAC}{digital-to-analogue converter}
\acro{DSP}{digital signal processor}
\acro{EuCAIF}{European Coalition for AI in Fundamental Physics}
\acro{EuCAIFCon}{EuCAIFCon 2025}
\acro{FPGA}{field programmable gate array}
\end{acronym}
\DeclareSymbolFont{usualmathcal}{OMS}{cmsy}{m}{n}
\DeclareSymbolFontAlphabet{\mathcal}{usualmathcal}

\DeclareSIUnit\flops{FLOPS}

\fancypagestyle{SPstyle}{
\fancyhf{}
\lhead{\colorbox{scipostdeepblue}{\bf \color{white} ~SciPost Physics Proceedings }}
\rhead{{\bf \color{scipostdeepblue} ~Submission }}
\renewcommand{\headrulewidth}{1pt}
\fancyfoot[C]{\textbf{\thepage}}
}

\begin{document}

\pagestyle{SPstyle}

\begin{center}{\Large \textbf{\color{scipostdeepblue}{
Applying reinforcement learning to optical cavity locking tasks: considerations on actor–critic architectures\\and real-time hardware implementation\\
}}}\end{center}

\begin{center}\textbf{
Mateusz~Bawaj\textsuperscript{$\star$1,2},
Andrea~Svizzeretto\textsuperscript{1,2}
}\end{center}

\begin{center}
{\bf 1} Universit\`{a} di Perugia, Italy
\\
{\bf 2} INFN, Sezione di Perugia, Italy
\\[\baselineskip]
$\star$ \href{mailto:mateusz.bawaj@unipg.it}{\small mateusz.bawaj@unipg.it}
\end{center}

\definecolor{palegray}{gray}{0.95}
\begin{center}
\colorbox{palegray}{
  \begin{tabular}{rr}
  \begin{minipage}{0.37\textwidth}
    \includegraphics[width=60mm]{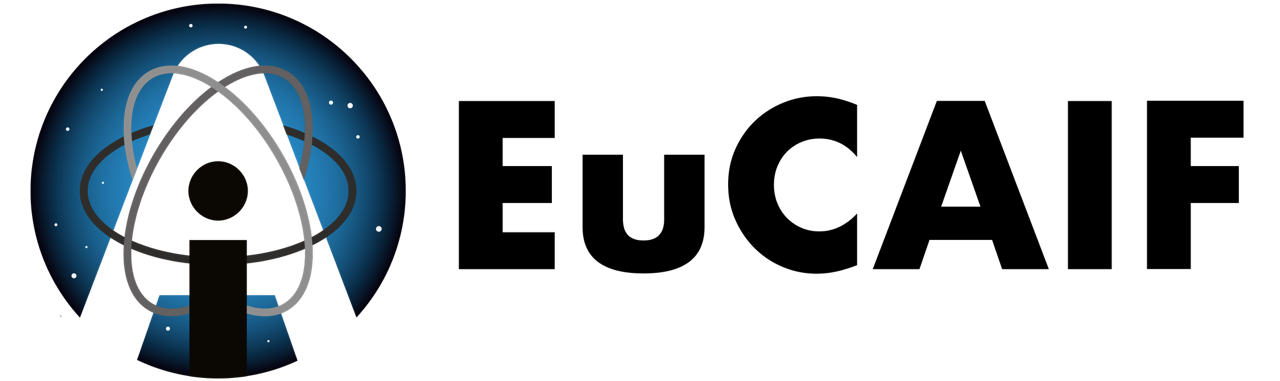}
  \end{minipage}
  &
  \begin{minipage}{0.5\textwidth}
    \vspace{5pt}
    \vspace{0.5\baselineskip} 
    \begin{center} \hspace{5pt}
    {\it The 2nd European AI for Fundamental \\Physics Conference (EuCAIFCon2025)} \\
    {\it Cagliari, Sardinia, 16-20 June 2025
    }
    \vspace{0.5\baselineskip} 
    \vspace{5pt}
    \end{center}
    
  \end{minipage}
\end{tabular}
}
\end{center}

\section*{\color{scipostdeepblue}{Abstract}}
\textbf{\boldmath{
This proceedings contains our considerations made during and after fruitful discussions held at EuCAIFCon 2025. We explore the use of \acl{DRL} for autonomous locking of Fabry–Perot optical cavities in non-linear regimes, with relevance to gravitational-wave detectors. A custom Gymnasium environment with a time-domain simulator enabled training of agents such as \acl{DDPG}, achieving reliable lock acquisition for both low- and high-finesse cavities, including Virgo-like parameters. We also discuss possible improvements with \acl{TD3}, \acl{SAC} and meta-reinforcement learning, as well as strategies for low-latency execution and off-line policy updates to address hardware limitations. These studies lay the groundwork for future deployment of \acl{RL}-based control in real optical setups.
}}
\vspace{\baselineskip}

\noindent\textcolor{white!90!black}{%
\fbox{\parbox{0.975\linewidth}{%
\textcolor{white!40!black}{\begin{tabular}{lr}%
  \begin{minipage}{0.6\textwidth}%
    {\small Copyright attribution to authors. \newline
    This work is a submission to SciPost Phys. Proc. \newline
    License information to appear upon publication. \newline
    Publication information to appear upon publication.}
  \end{minipage} & \begin{minipage}{0.4\textwidth}
    {\small Received Date \newline Accepted Date \newline Published Date}%
  \end{minipage}
\end{tabular}}
}}
}
\fancypagestyle{firstpage}{%
  \fancyhf{} 
  \fancyfoot[L]{\scriptsize Copyright 2025 CERN for the benefit of the ATLAS Collaboration. CC-BY-4.0 license.}
  \renewcommand{\headrulewidth}{0pt}
  \renewcommand{\footrulewidth}{0.4pt}
}

\thispagestyle{firstpage} 



\section{Introduction}
\label{sec:intro}
Experiments on \ac{GW} are not an exception to the worldwide trend of \ac{AI} evolution. Current experiments like LIGO, Virgo and KAGRA develop and implement new techniques, mostly focused on data elaboration~\cite{Cuoco_2025}. The next-generation detector, the Einstein Telescope~\cite{abac2025scienceeinsteintelescope}, will benefit from \ac{ML}-accelerated design optimization~\cite{Dorigo_2023} and generative AI-aided exploration~\cite{Krenn_2025}. We believe that our effort to involve \ac{ML} in control strategies for experimental setups is the right direction. In these proceedings, we briefly recall our results and describe new ideas born during the \ac{EuCAIFCon}.

We attempt a training of an RL agent in a custom Gymnasium~\cite{towers2024gymnasium} environment. The goal is to develop a strategy to autonomously lock a \ac{FP} optical cavity operating in a non-linear regime using a deep \ac{RL} agent, for future applications in \ac{GW} detectors~\cite{svizzeretto}. The work began with the design and validation of a highly accurate time-domain simulator that models cavity dynamics under varying mirror velocities, including cavity ring-down effects at high finesse. This simulator is embedded into a custom Gymnasium environment, enabling the \ac{RL} agent to interact with realistic cavity conditions.

The core control task, locking the cavity, is achieved by training a \ac{DDPG} agent~\cite{lillicrap2019continuouscontroldeepreinforcement}, which continuously adjusts mirror positions to reach and maintain resonance. We engineered a reward function based on cavity power and the \ac{PDH} error signal to guide the learning process~\cite{svizzeretto}. A carefully tuned reward function prevents the agent from learning oscillatory or suboptimal behaviours.

The agent is tested on both low- and high-finesse cavities, including those matching the parameters of the Virgo interferometer, demonstrating reliable lock acquisition even under strong nonlinearity. This work is currently being expanded to address the Sim2Real challenge by analysing delays and random effects and modifying the environment accordingly, with promising results in noisy scenarios. Partial observability and domain randomization are reported in a separate proceedings contribution. This work lays the foundation for future deployment of RL-driven control in real optical setups.

\section{Current status}

\begin{figure}
    \centering
    \includegraphics[width=0.9\linewidth]{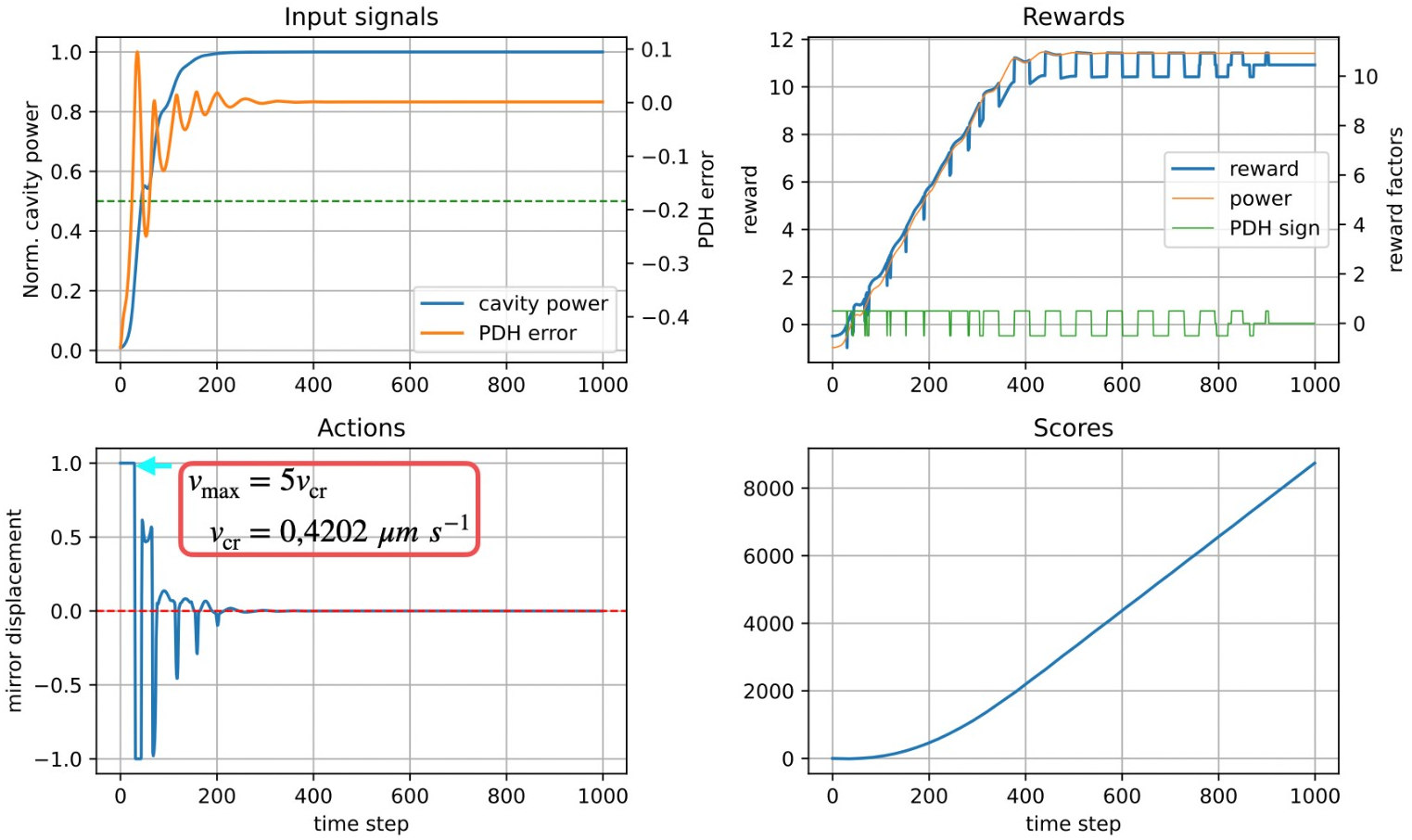}
    \caption{Performance of a standard \ac{DDPG} agent, trained for \SI{100}{\kilo{}} time-steps on a simulated cavity with parameters equivalent to the Virgo arm cavity. In this run, it takes approximately \num{200} steps to acquire the lock which is around \SI{10}{\milli\second}.}
    \label{fig:RLagent_lock}
\end{figure}

Trained agent is able to lock a cavity at the first resonance. See an example of a locking run in Fig.~\ref{fig:RLagent_lock}. The sequence starts at a random cavity length out of resonance. At the beginning, the agent changes the cavity length with the maximal allowed speed and when it finds the resonance, it modifies mirror movements accordingly to its policy. We would like to stress that the environment used for this training and demonstration does not simulate the finite force exerted by real actuators; thus, it does not simulate the inertia of the suspended mirror.

In such an environment, the agent training lasts from two to three minutes. At the first glance, it looks quick; however, we noticed that the training time does not change significantly when executed on two distinctly different hardware platforms. On both platforms, the environment calculates with \SI{32}{\bit} floating point precisions. The first platform is an M1 chip (8 core CPU, 8 core GPU, RAM \SI{8}{\giga\byte}) reaching realistically \SI{2}{\tera\flops}~\cite{hübner2025applevsorangesevaluating} while the second is a CUDA accelerated NVIDIA RTX A2000 GPU (\SI{4}{\giga\byte} VRAM) reaching almost \SI{6.5}{\tera\flops} (considering 80\% of nominal computing power). Clearly, the problem lays in the environment implementation, which does not benefit from parallel execution. Such slow training does not allow efficient domain randomization and further obstructs meta-learning training. We are going to address these issues in the next software release, together with optimization of the most time-consuming procedures using Numba~\cite{Lam_2015}.

\section{Improvements to the actor-critic architecture}
\label{sec:architecture}
Despite the \ac{DDPG} algorithm giving us the best results during training, there are more advanced successors worth our attention: \ac{TD3} \cite{fujimoto2018addressingfunctionapproximationerror}, known for its more accurate policy updates, and \ac{SAC}, which enables better exploration of the action space. In many tasks, both outperform \ac{DDPG} in providing more stable policies thanks to the use of two critic networks~\cite{fujimoto2018addressingfunctionapproximationerror,haarnoja2018softactorcriticoffpolicymaximum,Evans_2023}. The evolution of \ac{DDPG} was mainly discussed during the meeting; however, other, mostly application-focused studies~\cite{shengren2022performancecomparisondeeprl,yao2024performancecomparisondeeprl}, reveal that there is no clear winner between them and indicate \ac{PPO}~\cite{schulman2017proximalpolicyoptimizationalgorithms} as the best-performing \ac{DRL} algorithm in certain tasks. Discussions during \ac{EuCAIFCon} and the literature review motivate us to carry out thorough testing of all candidates in our environment.

Further, we discussed a possible evolution of our system toward \ac{Meta-RL}~\cite{Beck_2025}. This approach would shift the \ac{DRL} algorithm design problem to another level. Instead of manually tuning the algorithm, we would incorporate the tuning process into training in order to achieve better generalization, adaptability and efficiency across a wide spectrum of tasks.
At the moment, we do not have many diverse tasks to justify a continuous optimization of the policy leveraging meta-learning. However, modifying the environment in the future in order to support parallelization will enable the exploitation of the more robust generalization approach offered by meta-learning, compared to domain randomization.
Summarizing possible strategies, we think that currently \ac{Meta-RL} training is computationally too heavy for us and that the risk of producing algorithms unstable in implementation is too high. We therefore concluded that our current system will benefit from a temporal unit (\ac{RNN} or \ac{GRU}). Adding such a layer helps extract information from order-sensitive sequential data. We will focus on this development first.

\section{Low-latency execution}
\label{sec:hardware}
In feedback \acp{TDS}, the key aspect is the reaction time of the controller. Relevant contributions in digital controllers come from: sensor, \ac{ADC}, computing, \ac{DAC} and actuator propagation time. Our system differs from a classic \ac{DSP} only in the computing algorithm. Therefore, we will focus only on the \ac{RL}-agent inference computing time, known bottlenecks and possible techniques to improve it. As a reference for the data processing delay, we take two systems used to control the Virgo experiment~\cite{VirgoDAQ_2008}: the DSP for super-attenuator control, which guarantees a \SI{3,125}{\micro\second} delay, and the real-time DAQ, which guarantees \SI{100}{\micro\second}. Neither of the two systems is ready to accelerate \ac{ML}-agent inference. Therefore, our current setup uses a Jetson Nano, which computes \ac{DDPG} actor inference in \SI[separate-uncertainty=true,multi-part-units=single]{1.2(0.87)}{\milli\second}. The same chip is responsible for communication with the multichannel \ac{ADC} (ADS1256) and the \ac{DAC} (DAC8532). The bottleneck in this case is the actuator, which can be updated at a maximum rate of \SI{200}{\hertz}.

Discussions inspired by the work described in~\cite{Kaiser_2024} brought us subsequent ideas to address the above issues. We can apply two techniques to mitigate the low-frequency \ac{DAC} update limitation. The simplest solution to describe is to change the hardware. Our constraint on fast sampling and fast inference is naturally associated with the implementation of the agent into \acl{FPGA}-based electronics. The high-level synthesis language for machine learning HLS4ML~\cite{Duarte:2018ite} will be a natural choice for applying this solution.
However, an alternative strategy has attracted our interest. We can lower the \ac{RL}-agent action rate to a level sustainable by the hardware while keeping the \ac{ADC} acquisition rate high. According to experience from other applications, this could help the agent to forecast more accurate actions. In fact, we have noticed that an excessive sampling rate of our simulated environment does not bring benefits to the training.

The scarce resources of the Jetson Nano also make in-system policy optimization challenging. We find the possibility of off-line policy updates encouraging. In this case, the on-line agent is pre-trained and left running in the system. Periodically, data acquired during agent operation is transferred to an external computer, which performs policy optimization. The on-line system must then stop, reload the updated policy and resume operation. We plan to implement off-line policy updates as soon as we transfer our system to the real setup.

\section{Conclusion}
During \ac{EuCAIFCon} we reviewed our progress on applying reinforcement learning to the Fabry–Perot cavity locking problem and identified promising directions for further work. \ac{DDPG} agents trained in a custom Gymnasium environment demonstrated reliable lock acquisition and discussions highlighted possible improvements with \ac{TD3}, \ac{SAC} and meta-reinforcement learning. We also addressed the challenges of low-latency execution on embedded hardware and proposed mitigation strategies, including \ac{FPGA} implementation and off-line policy updates. These considerations strengthen the foundation for transferring our approach from simulation to real optical setups, with the long-term goal of integrating \ac{RL}-driven control into gravitational-wave experiments.

\section*{Acknowledgements}
We would like to express our sincere gratitude to Verena Kain and Michael Schenk from CERN for their fruitful discussions and valuable advice on alternatives to \ac{DDPG} in the context of our real-time control application. Their insights have been highly beneficial in shaping the direction of this work.
We would also like to thank Andrea Santamaria Garcia from the University of Liverpool for insightful discussions and valuable advice regarding online policy updates in our system.

\paragraph{Funding information}
Research is supported by Italian Ministry of University and Research (MUR) through the program “Dipartimenti di Eccellenza 2023-2027” (Grant SUPER-C), Italy.

\bibliography{bibliography.bib}

\end{document}